\documentclass[10pt]{article}
\usepackage{graphicx}
\usepackage{amsfonts}
\usepackage{amsmath}
\usepackage{amssymb}
\usepackage{mathrsfs} 
\usepackage{lipsum}
\usepackage{color}
\usepackage{siunitx}
\usepackage{authblk}
\usepackage[margin=2cm]{geometry}
\usepackage{mathtools, cuted}

\usepackage[T1]{fontenc}
\usepackage{graphicx}
\usepackage{hyperref}

\usepackage{bm}

\begin{document}

\vspace{10pt}

\title{The Life of a Surface Bubble}

\author{Jonas Miguet$^{1}$, Florence Rouyer$^{2}$ and Emmanuelle Rio$^{3}$\\
\small{$^{1}$ \quad TIPS C.P.165/67, Université Libre de Bruxelles, Av. F. Roosevelt 50, 1050 Brussels, Belgium; jonas.miguet@ulb.ac.be\\
$^{2}$ \quad Lab Navier, Univ Gustave Eiffel, Ecole des Ponts, CNRS, Marne-la-Vallée, France
; florence.rouyer@univ-eiffel.fr\\
$^{3}$ \quad Université Paris-Saclay, CNRS, Laboratoire de Physique des Solides, 91405, Orsay, France.; emmanuelle.rio@universite-paris-saclay.fr}}
\date{\today}

        \maketitle
           \begin{abstract}
Surface bubbles are present in many industrial processes and in nature, as well as in CO$_2$ beverage. They have motivated many theoretical, numerical and experimental works. This paper presents the current knowledge on the physics of surface bubbles lifetime and shows the diversity of mechanisms at play that depend on the  properties of the bath, the interfaces and the ambient air.
 In particular, we explore the role of drainage and evaporation on film thinning. 
 We highlight the existence of two different scenarios depending on whether the film cap ruptures at large or small thickness compared to the thickness at which van der Waals interaction come in to play.
           \end{abstract}


\section{Introduction}

Bubbles have attracted much attention in the past for several reasons.
First, their ephemeral nature commonly awakes children's interest and amusement. Their visual appeal has raised interest in painting \cite{behroozi2008soap}, in graphism \cite{huang2020chemomechanical} or in living art.
Nevertheless, these are not the only good reasons to study these rich physical objects (see for example \cite{de2013capillarity}). The more specific case of surface bubbles has been the focus of works conducted by various scientific communitie, due to their interaction with the environment. Indeed, 
a system in which a liquid is overlain by a gas and where surface bubbles are susceptible to appear is affected by these relatively small objects at the scale of the entire system. 
For instance, earth being covered by more than 70\% with oceans, researchers\cite{blanchard1963electrification,de2011production,veron2015ocean} found that bubbles-mediated production of sea-salt aerosols is a major natural source of cloud condensation nuclei.
In turn, the clouds affect the climate through their radiative properties, among other things. 
Bourouiba and Bush \cite{bourouiba2012drops} recently reviewed other examples of bubbles found in nature.
These include bubbles produced by gas exsolution in magmatic chambers \cite{gonnermann2007fluid}, local spreading of toxins through aerosolisation for instance during a Red tide event \cite{woodcock1948note,blanchard1989ejection} or even catastrophic release of oversaturated gas as happened in 1984 (Lake Monoun) and in 1986 (Lake Nyos) in Cameroon \cite{evans1996knowledge}.
Bubbles are also relevant in the glass industry where they favour homogenization of the melt but also constitute a major source of defects \cite{bolore2018spatial} and of energy loss as the formation of foam at the surface of the melt screens the heat coming from above. 
As the primary building block of a foam, the study of bubbles and their stability at the surface gets even broader, as was recently reviewed recently by Suja \textit{et al.}  \cite{suja2020single}. 
To take an image from the daily life, a beer drinker
may prefer a beer with some foam above it: drink a foam means drink a beer in Belgium.
A champagne consumer would probably not trust a foamed beverage: drink bubbles means consume sparkling wine in France. 
Beer drinkers and champagne consumers both love bubbles \cite{vega2017some,Liger-Belair2012}, but not in the same way.

\begin{figure}[htbp]
  \centering
    \includegraphics[width=10.5 cm]{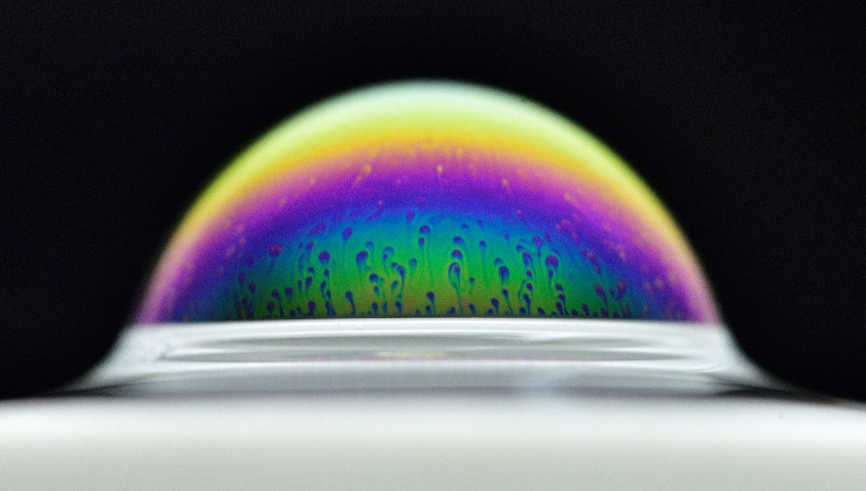}
  \caption{Artistic photography by Serge Guichard of a centimetric surface bubble during thinning. Marginal regeneration patches can be observed at the surface of a bubble. The bubble is stabilised by TTAB at a concentration of 0.5 times the critical micelle concentration (cmc). }
  \label{fig:BulleColoree}
\end{figure}

The entire life of a bubble covers many steps that we now shortly review to define the scope of this review.
Be it by nucleation in a supersaturated medium (exsolution), by drop or snow impact on a surface or by gas entrapment upon wave breaking, a bubble first needs to be born. 
Then, once the bubble is entrapped in a bulk liquid, the density difference fosters a movement that eventually brings the bubble to the surface. 
A typical distance for a rising bubble to reach its terminal velocity is one centimeter \cite{zawala2011influence,atasi2020lifetime}. 
During this process, the shear stresses induced by the flow may accumulate surfactants at the bottom of the bubble and affect the bubble shape and rising velocity. 
Once the bubble gets close to the surface,  they start interacting because the surface needs to be deformed and curved for the bubble to protrude. 
This step can lead to rebounds as shown for example in Zawala \textit{et al.} \cite{zawala2011influence}. 
When the bubble protrudes, a thin film is formed and it eventually adopts a metastable shape after typically 10-100 ms. 
The liquid film then undergoes thinning under the combined action of capillary and/or gravitational drainage,  evaporation and ultimately disjoining pressure. 
Eventually, a hole nucleates and invades the film, leading to the bursting of the bubble. 
This last step can trigger the formation of aerosols through three different processes: film drops due to the so-called "disgregation" of the thin film during rupture \cite{villermaux2020fragmentation}, jet drops due to the Worthington jet \cite{worthington1897v} and formation of daughter bubbles \cite{Bird2010}, themselves susceptible to undergo these different processes. 
Film drops and daughter bubbles formation depend upon the thickness of the thin film at burst \cite{Lhuissier2012,Bird2010}, itself conditioned by the thinning rate and lifetime of the bubble, thus the interest to study the whole process. 

In these problems, both gravitational and capillary effects must \textit{a priori} be taken into account. 
A key dimensionless quantity is therefore the Bond number \cite{hager2012wilfrid} - sometimes referred to as the Eötvös number - defined as :

\begin{equation}
Bo=\frac{\rho_{\text{liq}}gL^{2}}{\gamma},
\label{eq:BondNumber}
\end{equation}

\noindent where $\rho_{\text{liq}}$ [kg.m$^{-3}$] is the density of the liquid bath (considered much higher than that of the surrounding gas), g [m.s$^{-2}$] the gravitational acceleration and $\gamma$ [N.m$^{-1}$] the surface tension between the liquid and the gas. 
As we will see in the following, $L$ [m] is a geometric parameter that accounts for the phenomenon under consideration. Some authors use $L=R_\text{B}=\left(\frac{3V}{4\pi}\right)^{1/3}$, where V is the volume of the bubble, in part because $V$ is an adjustable parameter in numerical models. $R_\text{B}$ is then the radius of the equivalent sphere containing the gas volume within the bubble. Another relevant length is $R_\text{film}$, the radius of curvature of the emerged thin film (for large enough bubbles). It gives the magnitude of the surface tension-induced forces acting at the foot of the bubble cap and is easily accessible experimentally.
Quite naturally, the Bond number can be rewritten in terms of the capillary length, $\ell_{\text{c}}=\sqrt{\frac{\rho_{\text{liq}}g}{\gamma}}$, that appears in problems where gravity and capillarity dominate: $Bo=\frac{L^{2}}{\ell_{\text{c}}^{2}}$.

The scope of this review will focus on the life of bubbles at the surface of a liquid bath, \textit{i.e.} the shape, thinning and lifetime of surface bubbles, meaning we will consider 
bubbles with lifetimes typically of the order of 1-1000 seconds much larger than the time needed for the bubble shape to stabilise and for the hole to nucleate.
In this framework, we will consider the shape of the bubble at the surface in section \ref{section:shape}, the dynamic thinning of the thin film in sections \ref{section:thinning} and \ref{section:evaporation} and the consequences on bubbles lifetime in section \ref{section:lifetime}. 

\section{Shape of a surface bubble}
\label{section:shape}
Once a bubble has reached an interface, it adopts a static shape that is governed by the balance between the surface tension and buoyancy effects. 
In this case, the relevant length scale for the forces is $R_{\text{film}}$, defined as the radius of curvature of the spherical cap, both in the case of relatively small \cite{howell1999draining} and large \cite{Teixeira2015} bubbles. 
The corresponding Bond number is therefore defined as 
\begin{equation}
Bo_{\text{film}}=\frac{\rho_{\text{liq}}g R_{\text{film}}^{2}}{\gamma}.
\end{equation}

\begin{figure}[htbp]
  \centering
    \includegraphics[width=10.5 cm]{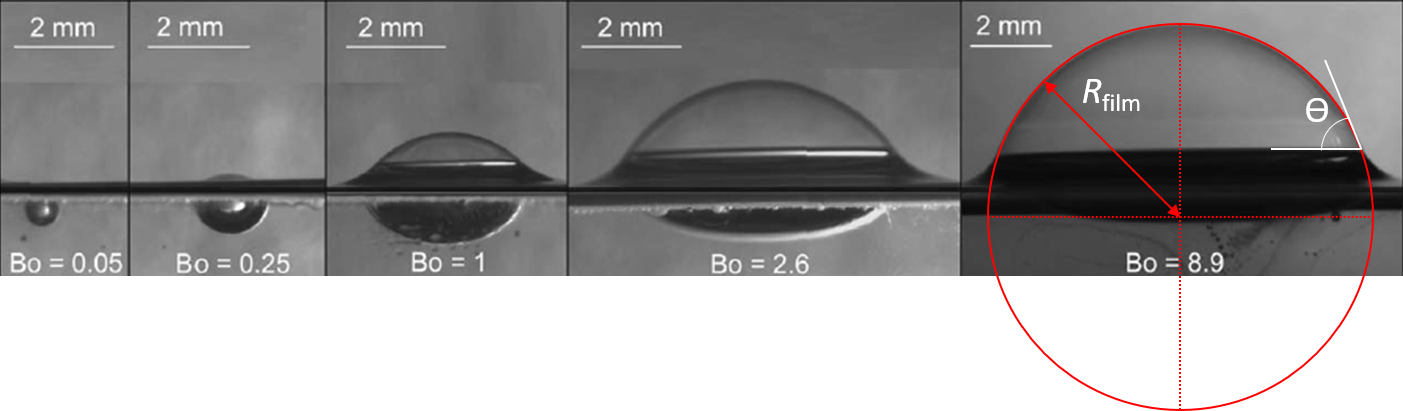}
  \caption{Adapted from Nguyen \textit{et al.}\cite{nguyen2013film}. Pictures of surface bubbles of varying Bond numbers.}
  \label{fig:fig1}
\end{figure}

\noindent Surface bubbles with varying Bond numbers are represented in Figure \ref{fig:fig1}. 
Two limit cases can be easily identified: 
if $Bo<<1$ surface tension dominates the system, buoyancy is not strong enough to deform the interface and  the bubble remains spherical and stands below the surface. 
For $Bo>>1$, the emerged spherical cap tends to a perfect hemisphere, the contact angle between the cap and the meniscus is equal to 90° and the bottom part of the bubble, below the free surface, is essentially flat  because hydrostatic pressure then compensates the over-pressure due to the curvature of the cap. 
Surface bubbles in natural or industrial processes often have intermediary sizes and the full resolution of the problem then requires to solve the Young-Laplace equation for the interface. Assuming axisymmetric geometry since surfaces (or, equivalently, free energy) must be minimum, three distinct zones have to be considered separately : 
the spherical cap made of a thin liquid film with two interfaces, above the meniscus, generally modeled as a weightless surface with a surface tension of $2\gamma$, so that the pressure jump across the interfaces writes $\frac{4\gamma}{R_{\text{film}}}$ as is the case in 3D; 
the immersed part of the bubble up to the top of the meniscus
, where the curvature of the interface and the hydrostatic pressure have to be taken into account ; and the outer side of the meniscus, where the curvature is balanced by the hydrostatic pressure and that is connected to the original bath height at infinity. 
No general analytical solution for the entire shape of the bubble can be inferred but numerous authors have tackled the problem \cite{nicolson1949interaction,Toba1959,princen1963shape,howell1999draining,pigeonneau2011low}. 
Most recently Teixeira \textit{et al.} \cite{Teixeira2015} derived a semi-analytical model valid in the limit of large Bond numbers (however providing convincing results down to approximately $Bo \sim 4$) and provided a comparison between experimental data, Surface Evolver simulations \cite{brakke1992surface} and to Howell's 2D and 3D models\cite{howell1999draining}. The better agreement obtained with 3D models suggests that the curvature in the horizontal plane should be taken into account, which does not contradict the axisymmetry of the problem.

This analysis shows that the apparent contact angle $\theta$, \textit{i.e.} the angle between the horizontal direction 
and the tangent to the spherical cap at the connection with the meniscus is a monotonously increasing function of the Bond number (Figure \ref{fig:fig1}).  
This implies that the bigger the bubbles, the larger the area of the thin film that eventually drains and ruptures. 
Finally as shown by Cohen \textit{et al.}\cite{Cohen2017}, the weightless film assumption significantly fails at describing the shape of bubbles of radii beyond a typical length higher than 10 cm (corresponding to Bond numbers of order 10 000), beyond which the emerged cap flattens and departs from its spherical shape. 

Finally, some useful length scales can be derived to simplify the analysis of experiments. 
In the case of bubbles with Bond numbers high enough to feature a protruding thin film and yet sufficiently small ($0.1<Bo<1$), gravity does not significantly deform the cavity 
that  keeps a spherical shape with a radius $R_{\text{cavity}}$. 
In this case, the equilibrium of the inner pressure in the bubble requires that $R_{\text{cavity}} \sim \frac{R_{\text{film}}}{2}$ which can be conveniently used to assess the curvature of the cavity for instance to study the subsequent formation of the Worthington jet\cite{seon2017effervescence}. 
Another geometrical parameter, important to write mass conservation in a spherical geometry, is the ratio between the perimeter $P$ [m] of the circle delimited by the top of the meniscus and the surface $S$ of the spherical cap. 
In the case of small bubbles Lhuissier \textit{et al.} \cite{Lhuissier2012} have shown that this ratio can be written $\frac{P}{S}\sim \frac{2l_{\text{c}}}{R_{\text{film}}^{2}}$. 
With the use of a numerical resolution of the Young-Laplace equation for the shape of the bubble, the authors show that this approximation works up to surprisingly large Bond numbers ($Bo\approx 5$). 
Finally for larger bubbles (and as long as the weight of the film may be neglected), this ratio may be expressed as $\frac{P}{S}\sim \frac{1}{R_{\text{film}}}$ with an error of less than a factor 2 for $Bo\geq 10$.

\section{Drainage}
\label{section:thinning}

Let us now describe the state of the art concerning the liquid drainage that is the liquid flow, mostly from the top to the bottom of the bubble, leading to film thinning.  

\subsection{Axisymmetric bubbles}

In the previous paragraph it is demonstrated that the static shape of the bubble is  axisymmetric along a vertical axis, as a consequence of gravitational and capillary forces balance. 
It is thus reasonable at first stage to assume that drainage is also axisymmetric and that the liquid flows from top to bottom inside a film whose thickness is supposed uniform for a given height.

The equation of motion of the liquid is governed by the Navier-Stokes equation : 

\begin{equation} 
\label{eq:N.S.}
    \rho_{\text{liq}} (\frac{\partial \mathbf{v}}{\partial t}+\mathbf{v. \nabla v })=-\mathbf{\nabla }P +\eta \Delta \mathbf{v} + \rho_{\text{liq}} \mathbf{g}.
\end{equation}

\noindent This differential equation of the liquid velocity $\mathbf{v}$ can be solved numerically provided  boundary conditions are defined at the liquid interfaces. 
Coupled with mass conservation, this equation allows to predict the film thickness evolution over time.  
For certain cases analytical solutions can be found as will be exposed later.

The left hand terms of this equation correspond to the inertia of the fluid. The right hand terms correspond successively to pressure, viscous and gravitational forces. 
Different non dimensional numbers are used to compare the terms of this equation with each others. 
As will be shown, different scaling analysis are proposed in the literature depending on the nature of the interfaces, \textit{i.e.} boundary conditions (the presence of surfactants or not), because it changes the characteristic length of the viscous dissipation.

The pressure scales with the Laplace pressure $\gamma / R_{film}$ and viscous stress with $\eta v / l_v^2$ where $l_v$ is the characteristic length of viscous dissipation. 
For practical reasons, in most numerical works, the equation is made dimensionless using $R_B$ assuming $R_{\text{film}} \sim R_{\text{B}}$. 

\subsubsection{Viscous bubble from bare to not so bare interfaces}

It is common for the problem of bubbles at the surface of viscous fluids as for example molten glass, magma or polymeric fluids  to assume that the inertia of the liquid flow is negligible. Moreover, these fluids are supposed to have perfectly bare interfaces, \textit{i.e.} no surfactants nor impurities so that the interfaces are supposed stress free. Such hypothesis implies a plug flow of the liquid. The flow is thus extensional, \textit{i.e.} the viscous dissipation is due to the velocity variation along the flow. 

The non dimensional number, that compares inertia to viscous dissipation, is the Reynolds number and writes $Re=\rho_{\text{liq}}^2 g R_B^3 / \eta^2 $, assuming a motion driven by gravity and a length scale equal to the bubble size with a velocity scaling as $\rho g R_B^2/\eta$. Neglecting inertia implies $Re<<1$. 
One can note that for air bubble in water, this criterion sets a maximal bubble size to $10$ $\mu$m  whereas for viscous fluids ($\eta=1$ Pa.s) it sets it to $1$ mm.
Additionally, two viscous characteristic times can  be deduced from equation \eqref{eq:N.S.}, one is based on bubble buoyancy $\tau_{B} = \frac{\eta}{\rho_{\text{liq}} g R_B } $, and the other one is based on capillarity at the scale of non deformed bubble: $\tau_{Cb} = \frac{\eta R_B}{\gamma }$. Their ratio leads to a Bond number $Bo_B=\tau_{Cb}/\tau_{B}$ and $\tau_{B}$ is assumed to be relevant for large $Bo_B$ and $\tau_{Cb}$ for low $Bo_B$.

\begin{figure}[htbp]
  \centering
    \includegraphics[width=10.5 cm]{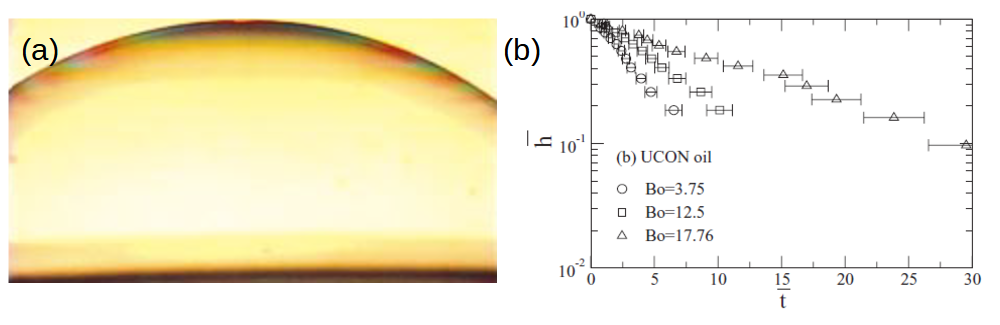}
  \caption{(a) View of a large bubble at the surface silicon oil, extracted from Debregeas et al. \cite{Debregeas1998}, (b) Normalized film thickness $\bar{h}=h/h_0$ as a function of normalized time $\bar{t}=t/\tau_g$ for different bubble sizes at the surface of Ucon Oil for different $Bo_B$ numbers corresponding to different buble sizes, extracted from Kocarkova et al. \cite{kovcarkova2013film}}
  \label{fig:figdrainviscous}
\end{figure}

Since the pioneer quantitative work on the subject from G. Degregeas, P.G. De Gennes and F. Brochard \cite{Debregeas1998} for large surface bubble, it is known that the film thickness $h$ decreases monotonously from bottom to top (Figure \ref{fig:figdrainviscous} (a)) and decays exponentially (Figure \ref{fig:figdrainviscous}(b)). One can write:

\begin{equation}
    h(t)=h_0 \exp(-a t/\tau)
\end{equation} 

\noindent where $\tau$ is a characteristic viscous time. The value of $a$ is equal to the normalized thinning rate of the film: $a = - (dh/dt) (\tau/h_0) $. Their experiments with silicon oil and quite large bubble ($Bo>1$) give $a_g\sim 0.21$ considering $\tau=\tau_B$ which is consistent even though lower than their analytical model that proposed $a_g=1.25$ (see \cite{kovcarkova2013film}).

Numerical simulations, based on Stokes equation (no inertia, $Re=0$), confirm the exponential decay of the film thickness \cite{pigeonneau2011low} and show a decay of $a$ with the Bond number $Bo_B$ which tends to $a_{g \infty}=0.34$, in the limit of large bubbles (large $Bo$)\cite{kovcarkova2013film,atasi2020lifetime}. This result implies that in normalized time, the smaller the bubble, the faster it drains.
Experiments \cite{kovcarkova2013film} with molten glass and various viscous fluids (Ucon, Castor oil)  provide $h(t)$ for different bubble sizes (Figure \ref{fig:figdrainviscous}(b)), the data are dispersed, even so the variations of $a$ with $Bo_B$ are consistent with previous simulations.
In a second set of experiments, Nguyen \textit{et al} \cite{nguyen2013film} have measured the time  $\Delta t$  between the moment at which bubble rising velocity becomes negligible and the one at which
the bubble spontaneously ruptures.
Then, assuming that this time interval  corresponds to the time required for the film to drain from an initial thickness ($h_0\sim 100$ $\mu $m) down to a final thickness ($h_f\sim 100$ nm) and assuming an exponential decay of $h(t)$, they are able to calculate the normalized thinning rate

\begin{equation}
    a=\ln(h_0/h_f)/(\Delta t/\tau). 
\end{equation}

Their results agree quite well with previous ones, and we estimate $a_{g \infty}=0.2$ once converting the normalization with the same parameters than above. They also proposed a scaling with $\tau_\text{c}$ and find $a_\text{c}= 0.05$  at low $Bo$.

In the limit of small bubbles, (low Bond number) Howell \cite{howell1999draining} studied the drainage of bubbles at the free surface using a lubrication model in 2D and pointed out that the film thickness decays according to an algebraic function of time.

\begin{equation}
    h(t)=h_0(1+ct)^{-2}, \text{where } c\propto\frac{\gamma^2 h_0^{1/2}}{\eta \rho g R_{\text{film}}^{7/2}}
\end{equation}

Finally, recent numerical simulations of viscous bubbles ($Re$=0.1) have modelled the thinning of the film on top of surface bubbles in presence of surfactants \cite{atasi2020lifetime}, thus the interfaces are not bare anymore. The simulations are able to model Marangoni stress induced by the inhomogeneity of surface tension due to surfactants concentration variation at the interfaces. They retrieve an exponential decay of the film thickness over time and a monotonous decay of the film thickness from the bottom to the top, for low concentration of insoluble surfactants or for very large bubbles ($Bo>1$). 
However they find an algebric decay at low $Bo$ and large surfactants concentration which reduces the drainage rate (long living bubbles). 
The presence and dynamics of surfactants may change the equilibrium shape of the bubble,  as previously observed for a bubble and a free surface with unequal surface tensions \cite{guemas2015}. 
Additionally, the presence of surface tension gradients not only rigidifies the interfaces but also affects the usual monotonous thickness decay from top to bottom.
Indeed, for intermediate bubble size and large surfactant concentration, a pinch (neck) appears at the base of the cap  (Figure \ref{fig:pinch}(a)). 

\subsubsection{Apparition of a pinch or a dimple}
\label{section:pinching}

\begin{figure}[!ht]
  \centering
    \includegraphics[width=1\linewidth]{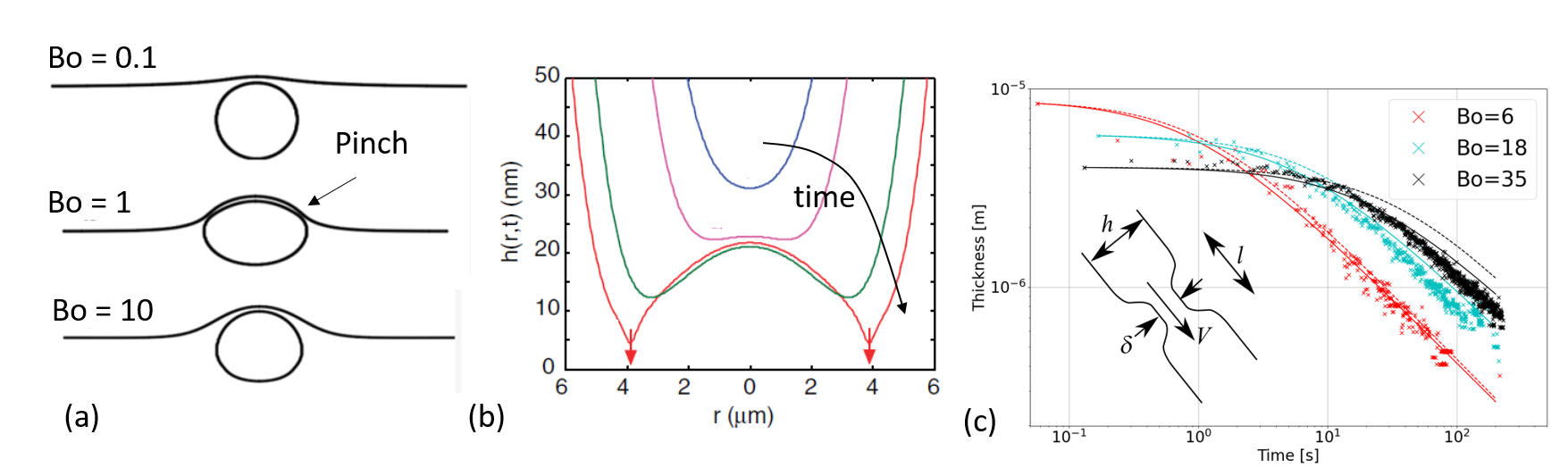}
  \caption{(a) Scheme adapted from reference \cite{atasi2020lifetime}. From top to bottom, shape of bubbles at increasing Bond numbers in presence of some surfactants. For intermediate Bond numbers, a pinch appears, which leads to a drastic increase of the bubble lifetime. (b) Prediction of the thickness profile of the film during a bubble-bubble approach. $r$ is the radial coordinate with its origin on the vertical axis of symmetry of the two approaching bubbles. A pinch appears and thins along time. Image adapted from reference \cite{Vakarelski2010}. (c) Comparison of the thickness of the film at the top of bubble with the model proposed by Lhuissier \cite{Lhuissier2012} (dashed lines) complemented by a gravity driven drainage (solid lines) for bubbles of different radius $R$. The inset represents a sketch of the pinch in the vicinity of the meniscus, close to the bath. Figure adapted from reference \cite{Miguet2019}. }
  \label{fig:pinch}
\end{figure}

Let us now concentrate on this pinch, which can appear close to the meniscus.
Equivalently, this pinch can be seen as a thicker zone or dimple at the top of the bubble.
As mentioned in the previous paragraph, the pinch indeed leads to a slowing down of the liquid flow, and to long living bubbles \cite{atasi2020lifetime}. 
As we will see in the following, this pinch can destabilise in small patches leading to a loss of axisymmetry.
Here we concentrate on the situations, in which an axisymmetric description holds.

This pinch was already mentioned and observed by Mysels in the 60' \cite{Mysels1959} in vertical films.
It is attributed to the capillary suction in the meniscus and
Aradian \textit{et al} provided a theoretical description of the pinching dynamics \cite{aradian2001marginal}. 
They proposed scaling laws predicting the velocity at which the pinch thins and enlarges along time.
Although this pinch has been observed in many different experimental studies (see for example \cite{Chan2011,Gros2020,Bhamla2017}), its dynamics has
never been compared to the theoretical prediction by Aradian. 
This is due to a fast destabilisation in many experimental studies, which will be discussed in the next section and to difficulties to measure the thickness with high enough accuracy. 

The presence of a surface tension gradient seems necessary for this pinch to appear. 
Indeed,
Howell and Stone \cite{howell2005absence} showed theoretically that no pinch can develop in absence of surface tension gradients.
This has been confirmed more recently by Atasi \textit{et al} \cite{atasi2020lifetime} with numerical simulations.

Chan and collaborators \cite{Chan2011} also described a pinching at the periphery or, equivalenty, a dimple at the center, when two bubbles approach at controlled velocity in an Atomic Force Microscope (AFM). 
They link the apparition of the dimple to a balance between the Laplace pressure, which tend to keep a round bubble on one side and the hydrodynamical and disjoining pressure, which tend to flatten the film on the other side.
The result leads to the evolution of the bubble profile along time plotted in Figure \ref{fig:pinch}(b).
In this figure, the axisymmetric profile exhibits a pinch, whose thickness decreases along time until the bubbles coalescence.
They obtain a very nice description of the experimental bubbles profiles obtained by interferometry.
As noticed in the review on bubbles and drops in AFM written by Wang\cite{Wang2015}, a direct consequence of these results is that a dimple is less expected to appear in small bubbles because they are less deformable. 
This in agreement with Atasi \textit{et al} \cite{atasi2020lifetime}, whose numerical solutions exhibit a range of Bond numbers, typically, between 0.3 and 3, in which a pinch can be observed (see typical shapes in Figure \ref{fig:pinch} (a)).

Lhuissier \textit{et al} \cite{Lhuissier2012} proposed a scaling law to take into account the presence of the pinch in the drainage dynamics.
The principal ingredient of their model is that the pinch limits the liquid flow driven by capillary pressure toward the bottom of the surface bubble.
The principal steps of the calculation are described in the following.
For small bubbles, the mass conservation in a spherical cap writes
 \begin{equation}
\frac{\partial h}{\partial t} + \frac{\ell_c}{R_{\text{Film}}^2} V h =0. \label{eq:LhuissierMassConservation}
 \end{equation}
\noindent where $V$ is the average fluid velocity, set by the capillary suction in the meniscus
and can be extracted by scaling the Stokes equation
\begin{equation}
    \eta \frac{V}{\delta^2} \sim \frac{\gamma}{R_{\text{film}} \ell},
    \label{eq:LhuissierStokes}
\end{equation}
\noindent in which $\delta$ and $\ell$ are respectively the thickness and the width of the pinch (insert figure \ref{fig:pinch}(c).
The thickness $\delta$ is expected to scale with $h$ while $\ell$ is set by a curvature matching between the curvature $\frac{h}{\ell^2}$ in the pinch and the curvature $\frac{1}{R_{\text{film}}}$ of the bubble.
Altogether, this leads to a prediction for the thinning law,
\begin{equation}
h(t) \sim \left(\frac{\eta}{\ell_c \gamma t}\right)^{2/3} R_{\text{film}}^{7/3}.
\label{eq:LhuissierScaling}
\end{equation}
Note that the fact that $\delta$ scales with $h$ comes from the fact that the pinch thinning is interrupted by the pinch destabilisation (marginal regeneration), which we will describe in the next paragraph. 
The model is thus axisymmetric even though it necessitates the existence of this destabilisation.

As we will show later (section \ref{section:evaporation}), this prediction only holds if evaporation can be neglected, \textit{i.e.} for reasonably thick films or in saturated atmosphere.
In these conditions, the scaling has been verified experimentally using various methods to measure the bubble thickness. 
Miguet \textit{et al} \cite{Miguet2019} used a spectrometer to benefit from the interferences and extract the thickness at the top of the bubble cap. 
Another classical measurement \cite{poulain2018,Poulain2018a, Lhuissier2012,Modini2013} consists in measuring the opening velocity of the cap when the bubble bursts. 
The Culick law \cite{Culick1960} then allows to link the film thickness to the opening velocity $v = \sqrt{\frac{2 \gamma}{\rho h}}$. 
For example, in Figure \ref{fig:pinch}.c, the thickness measured by interferometry at the top of the bubble is plotted as a function of time for different bubble sizes in a saturated environment. The dashed lines represent the integration of Eq. \ref{eq:LhuissierMassConservation} using the capillary driven drainage to calculate the velocity.
The solid lines represent a similar integration taking into account not only the capillary driven drainage but also the gravity driven one.
It is not surprising to see that the gravity must be taken into account to describe accurately the drainage of the bigger bubbles \cite{Miguet2019}. 

\subsection{Loss of axisymmetry}
\label{section:destabilisation}

\begin{figure}[!ht]
  \centering
    \includegraphics[width=1\linewidth]{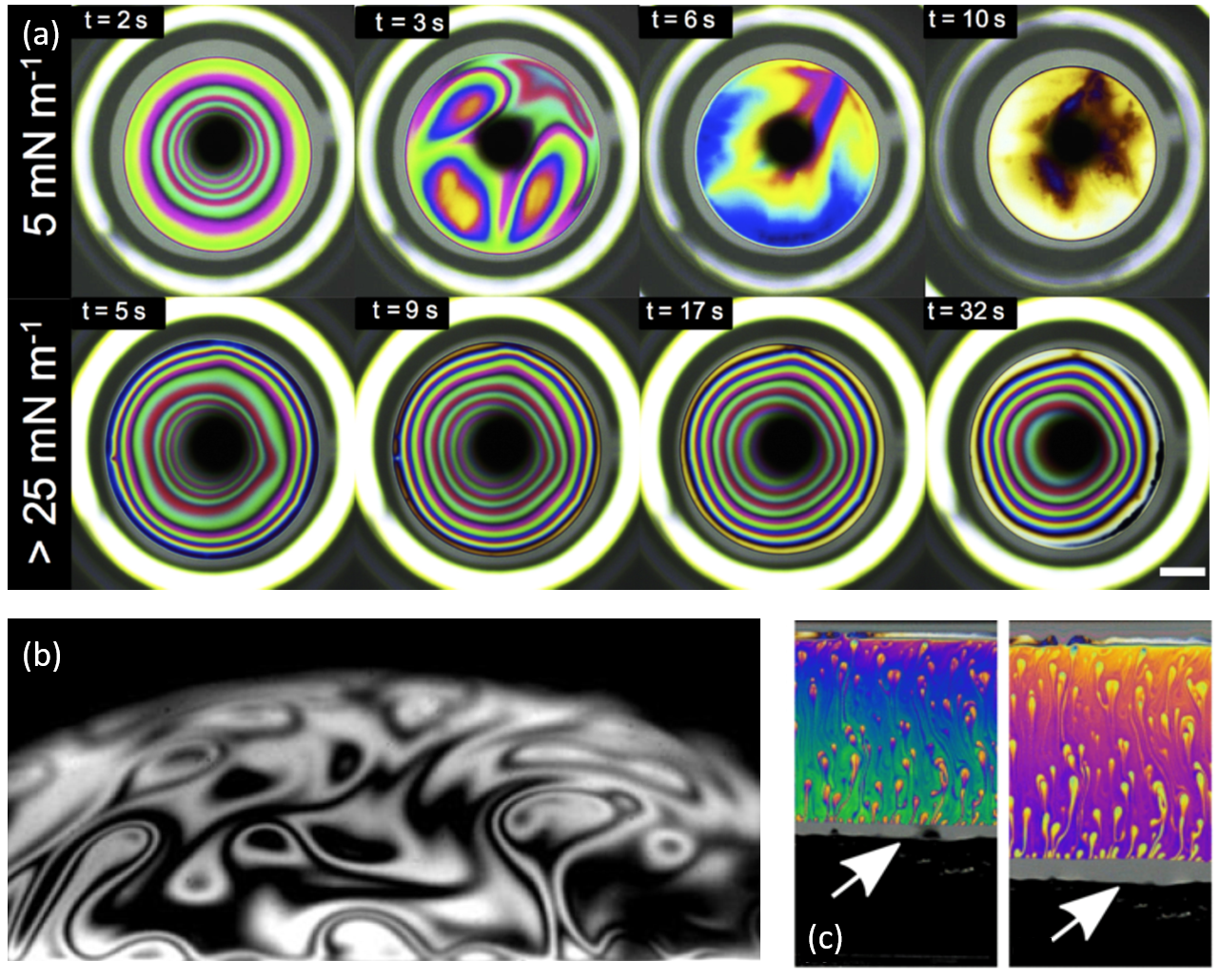}
  \caption{(a) Top view of a bubble thinning along time. The bubbles are stabilised by DPPC. The experiment allows to control accurately the surface pressure, which is set to 5 mN.m$^{-1}$ (top) or to 25 mN.m$^{-1}$. The former exhibits some loss of axisymmetry. Figure adapted from reference \cite{Bhamla2017} (b) View of the marginal regeneration patches in the meniscus of a surface bubble. Figure adapted from reference \cite{Lhuissier2012} (c) Observation of marginal regeneration patches rising at the surface of vertical foam film during its extension at a controlled velocity. The white arrow indicates the position of the frame used to extend the foam film. Figure adapted from reference \cite{shabalina2019}.}
  \label{fig:marginalregeneration}
\end{figure}

The pinch described in the preceding section happens to be unstable in certain conditions.
The axisymmetric thin zone at the bottom of the bubble can indeed destabilise in thin patches.
Their diameter is around a few hundreds of micrometers.
These thinner and therefore lighter zones tend to rise along the interface due to buoyancy because of their smaller apparent density \cite{Adami2014} and are reminiscent of the dynamic colored zones, which can be observed  in Figure \ref{fig:marginalregeneration} (b) and (c). 
Mysels chose to call this phenomenon \textit{marginal regeneration} because new thin zones enter the foam film and replace the aspiration of thicker zones inside the meniscus.
The regeneration mentioned by Mysels thus refers to a regeneration of the surface with thinner and more fragile zones.
This finally contributes to the global film thinning.
As noticed by Aradian \cite{aradian2001marginal}, Atasi \textit{et al} \cite{atasi2020lifetime} as well as by Chan \textit{et al} \cite{Chan2011}, one of the future challenge to understand fully bubble drainage is to better describe this loss of axi-symmetry and its impact on drainage. 

Let us first raise that this destabilisation necessitates $mobile$ enough interfaces.
The extreme opposite case is the presence of a solid surface.
Indeed, if the thin film is deposited on a solid surface,  the capillary suction by the meniscus is the same so that the pinching should occur and its dynamics should be described by Aradian \textit{et al} \cite{aradian2001marginal}.
Nevertheless, the presence of a solid happens to kill the instability \cite{Hack2018}.
This is actually the only situation, in which quantitative measurements have been performed.
Hack \textit{et al} \cite{Hack2018} measured the film profile around drops by Digital Holography Microscopy (DHM).
They enlighten the presence of the pinch and measure widening of the pinch position scaling with $t^{1/4}$ as predicted by Aradian \textit{et al} \cite{aradian2001marginal}.
Nevertheless, they observe no destabilisation.
This is confirmed by the experiments performed by Bhamla and collaborators, who observed this destabilisation for liquid/air interfaces (Figure \ref{fig:marginalregeneration}(a)) and not for solid/liquid ones \cite{Bhamla2017}. 

The mechanical resistance of the interfaces can thus prevent the appearance of marginal regeneration.
This is reminiscent of the observations by Mysels \cite{Mysels1959} showing that the marginal regeneration is observed when the films are stabilised by Sodium Dodecyl Sulfate (a standard surfactant) but disappears if a co-surfactant, such as dodecanol is added to the solution.
This co-surfactant is indeed known to allow the surface to sustain large surface tension gradients and thus to exhibit a large surface Gibbs elasticity.
Mysels chose to call such interfaces \textit{rigid} because of the absence of motion due to marginal regeneration and we can only guess that this rigidity has an effect similar to the solid surface mentioned above.
 Bahmla \textit{et al} \cite{Bhamla2017} have also observed that large  surface elasticity can prevent the development of the instability.
 They use surfaces covered by a phospholipid, the DPPC (dipalmitoylphosphatidylcholine) and they control the surface tension by using a Langmuir trough. 
 Then, they create a bubble at the liquid/air interface and visualize the drop profile thanks to the colors characteristic of the interferences.
 Figure \ref{fig:marginalregeneration} (a) shows such bubbles created at smaller (top) and higher (bottom) surface concentrations in DPPC.
 The axial symmetry is broken for small concentrations, which also corresponds to small surface elasticity, showing that the surface rigidity can prevent the pinch destabilisation.

Finally, in vertical soap films as well as in horizontal ones or in surface bubbles, the pinch instability and the subsequent marginal regeneration is expected for large enough characteristic sizes, in presence of standard surfactants \cite{shabalina2019,Gros2020,Miguet2020, Lhuissier2012} leading to low enough surface elasticity.
Mysels already proposed that the dynamics of these thin films is crucial to understand the foam film thinning since thin patches tend to rise and replace thicker ones. 
It has recently been shown that this \textit{sliding puzzle dynamics}, as Gros and collaborators have called this phenomenon, can indeed fully explain drainage in a film \cite{Seiwert2017} or in a surface bubble \cite{Miguet2020}.

Surprisingly we now have two mechanisms to describe the drainage of a thin film in presence of marginal regeneration: the model proposed by Lhuissier \textit{et al} \cite{Lhuissier2012}, in which the flow is determined by the friction in the pinch and the sliding puzzle dynamics, in which the thinning is due to the reorganization of the thin and thick zones.
Both are susceptible to describe very well the drainage in presence of marginal regeneration \cite{Miguet2019,Poulain2018a,Seiwert2017}.
We are currently investigating the sliding puzzle dynamics on surface bubbles.
Both phenomenon are actually coupled.
The pinch thins until a critical thickness at which the destabilisation occurs.
This thickness fixes the viscous dissipation in the film.
It also fixes the patch thickness and thus its efficiency to contribute to film drainage.
The film must find a working point in which both thinning velocities are compatible.

\section{Evaporation}
\label{section:evaporation}

\begin{figure}[!ht]
  \centering
    \includegraphics[width=1\linewidth]{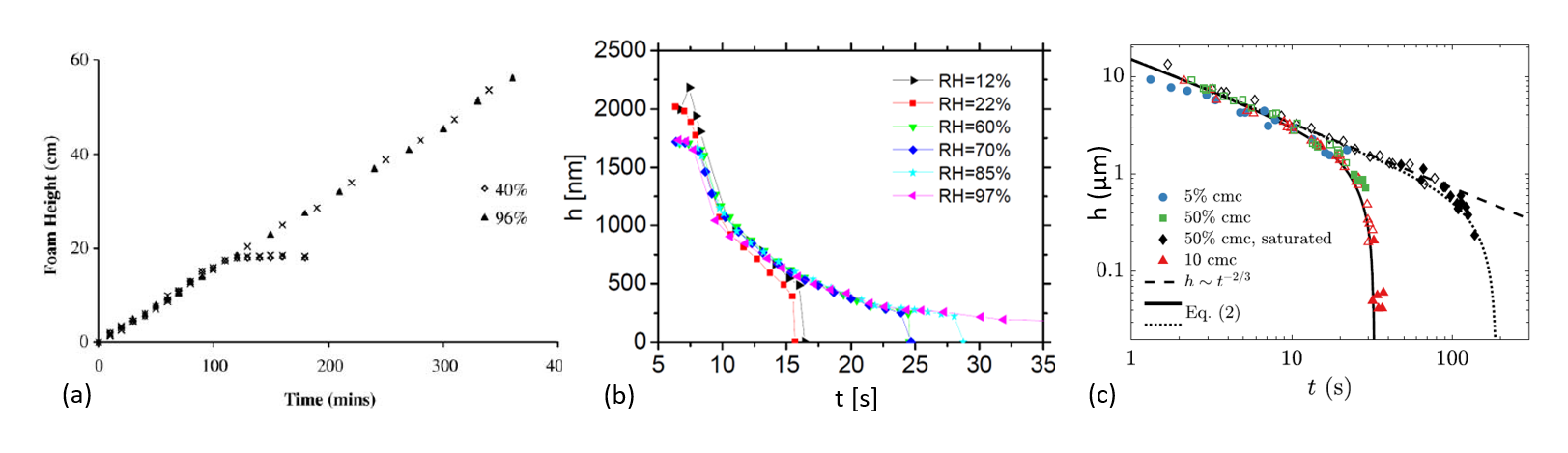}
  \caption{Representative experiments of thin films stability with controlled humidity. (a)From Li \textit{el al.}\cite{Li2010}. Foam height as a function of time in a cylindrical column which height is 55 cm between the liquid interface in the absence of foam and a humidity sensor. The bottom of the column is immersed in a 2.92 g.L$^{-1}$ SDS aqueous solution, where a constant bubbling produces the foam. The relative humidity at the top of the column is controlled. Cross symbols represent repeat of experiments. (b) From Champougny \textit{et al.} \cite{Champougny2018}. Thickness evolution 1 mm below the superior frontier of a thin film extracted at 0.5 mm.s$^{-1}$ from a 5 cmc TTAB solution. The h=0 nm points correspond to the instant when the film rupture is detected. (c) From Poulain \textit{et al.}\cite{poulain2018}. Film thickness evolution of surface bubbles with $R$=5.2 $\pm$ 0.4 mm in solutions of TTAB of concentration 0.05, 0.5 and 10 times de cmc. The unsaturated experiments are performed in ambient conditions \textit{ie} RH=26\% while the saturated ones are about RH=96\%. The continuous and dotted lines are the solutions of the integration of equation \ref{eq:PoulainMassConservation}, respectively with RH=26\% and RH=96\% The thickness measurements are operated through Taylor-Culick velocity measurements at burst. Filled symbols represent spontaneous burst events while open ones are obtained through manual bursts with a needle.}
  \label{fig:evaporation}
\end{figure}

Because it is not relevant in a number of thin film stability problems (foam or emulsion coalescence for instance), evaporation is a rather overlooked subject in literature. 
As was pointed out by Li \textit{et al.}\cite{Li2010}, the relevance of this parameter is however accessible in  the everyday life. If the reckless pouring of a carbonated beverage like a beer can lead, at least in a first time, to high foam to liquid ratio, it does not always end up in overflow. 
The rise of the foam indeed often ends just at the surface of the glass which is not only due to luck, but also because the destruction rate of the foam is higher there, because evaporation is higher than within the glass column. 
The experiment of these authors is represented in figure \ref{fig:evaporation}(a). 
A cylindrical column at the bottom of which a soapy liquid is constantly bubbled and at the top of which the relative humidity is controlled is used to measure the height of the foam as a function of time. 
The results show that, in the case of low humidity, a constant height of equilibrium is reached while the foam continuously rises to the top of the column when the humidity is saturated. 
The explanation is that the rate of destruction of the produced foam is high enough to balance the rate of production when evaporation proceeds. 
On the contrary, when humidity is close to saturation and therefore evaporation eliminated, the foam destruction rate is much lower and the foam height can reach the limit of detection of the set-up.
This illustrates the leading order role of evaporation in surfactant-stabilised films.

Going one step further, Champougny \textit{et al.}\cite{champougny2018influence} measured the thickness evolution of planar thin films during the continuous extraction of a frame from a soapy solution in a humidity controlled environment. The data are presented in figure \ref{fig:evaporation}(b). 
The main conclusion is that the evaporation rate does not significantly influence the thinning rate of the film but determines to leading order the time (or, equivalently, the length) at which it bursts. 
A satisfactory explanation for such observations is then presented by Poulain \textit{et al.} \cite{Poulain2018a} on millimetric-sized surface bubbles. As can be seen when derivating equation \ref{eq:LhuissierScaling}, the thinning rate due to drainage $\frac{dh}{dt}$ is a monotonically decreasing function of time (in absolute value) so that the thinning due to drainage is expected to become very slow along time. 
As a first approach, the tinning rate due to evaporation $J$ [m.$s^{-1}$] is constant and can be estimated from diffusive models written for drops \cite{fuchs1959evaporation}, giving typical values or 1 to 10 nm.s$^{-1}$ respectively for centimetric and milimetric bubbles. 
By equalizing the thinning rate due to drainage with the one due to evaporation, one can find an expression for a critical thickness, above which the thinning is dominated by drainage and below which evaporation becomes more important. 
This thickness is approximately 1 $\mu$m for millimetric bubbles.
To go one step further,
mass conservation  needs to be rewritten to take this contribution into account and reads : 
\begin{equation}
\frac{\partial h}{\partial t} + \frac{\ell_c}{R_\text{film}^2} V h + J=0, 
\label{eq:PoulainMassConservation}
 \end{equation}
\noindent This law is verified on figure \ref{fig:evaporation}(c). 
These results demonstrate the importance of evaporation on the thinning dynamics.
 Additional results concerning the impact of evaporation on the thinning of surface bubbles were obtained by Miguet \textit{et al.}\cite{Miguet2019} and confirm the approach of Poulain \textit{et al.}\cite{Poulain2018a}. 
In particular, they show the importance of gravitational convection due to water vapor being lighter than dry air. 
This convection is shown to determine the water vapor concentration field and therefore the evaporation rate above the bubbles for a circular bath with a radius of 2 cm. A typical value for the evaporation rate in this case is a few tenths of nanometers per second, independently of the bubble size. 
This is expected to be true as soon as the characteristic size of the system is above a few millimeters. This remark thus probably holds for carbonated beverages and more generally for most natural and artificial systems where surface bubbles appear. 

An important remark concerns the general shape of the curves depicting thickness evolution. 
It turns out that exponential laws have been found to match experimental data very well in the case of TTAB (Tetradecyl Trimethyl Ammonium Bromide) bubbles far from saturation conditions\cite{Champougny2016}. The results shown in \cite{poulain2018,Miguet2019} however show that the power law behaviour performs clearly better in saturated conditions.
The addition of the evaporation rate to a power law drainage regime describes very well the data, even so they look like exponential decays.

It should also be noted that Poulain \textit{et al.} \cite{poulain2018} reported results for surface bubbles made of "dirty" or salt water with poor stability (generally less than 20 s) for which the temperature of the liquid bath was systematically controlled thus tuning the evaporation rate. They make the counter-intuitive observation that bubbles submitted to higher evaporation rates feature thicker films. The results are rationalised by invoking thermal Marangoni backflow from the hot bath to the cold film, where surface tension is then decreased. A similar behaviour was reported for flat films made of melted glass  \cite{pigeonneau2012stability}.

The stabilisation effect of evaporation is mentioned in the review by Suja \textit{et al.} \cite{suja2020single} and supported by Lorenceau \textit{et al.} \cite{Lorenceau2020} notably with the remark that bubbles formed in an ethanol/water binary solution are completely unstable if the ambient atmosphere is saturated in ethanol while they live generally tens of seconds when evaporation is allowed to proceed. 

The condition(s) that determine the path from the stabilising effect of evaporation at the beginning to its destabilising effect at late stage remain(s) unknown.
Nevertheless, evaporation is crucial for the determination of the bubbles lifetime as will be shown in the following.

\section{Bubble lifetime}
\label{section:lifetime}

Who has never observed at the surface of a puddle under the rain one bubble that bursts instantly and another one that stays for more than 10 seconds? Same can be said for bubbles at a surface of sodas or beverage with alcohol, and fruit juice.

At first sight, these liquids have no stabilising molecules in a sense that no surfactants are present in the solution to generate a positive disjoining pressure which balances van der Waals forces and thus prevents the interfaces of a thin films from merging \cite{israelachvili2011intermolecular}. 
Moreover, we highlight that such stabilising molecules would act at very short distance between both interfaces (typically 100-10 nm), whereas it is unlikely, based on previous paragraphs, that the liquid film on top of a millimetric bubble has reached such low thickness value in less than a second.  

\begin{figure}[!ht]
  \centering
    \includegraphics[width=1\linewidth]{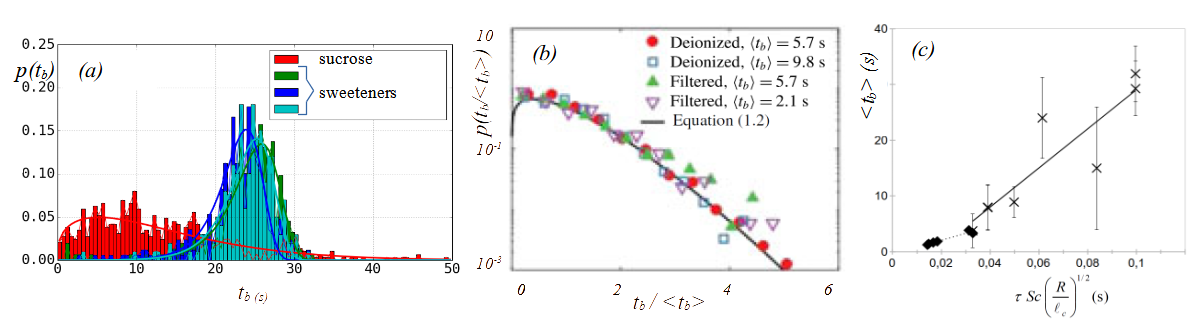}
  \caption{(a) Probability density function of lifetimes of bubbles at the surface of aqueous solutions of sucrose and sweeteners, extracted from Miguet et al. \cite{Miguet2020}  (b) Normalized probability density function of lifetimes of bubbles at the surface of water baths of different quality, extracted from Poulain et al. \cite{poulain2018} (c) Average lifetime of bubbles as a functions of $\tau Sc (R/l_c)^{1/2}$ at the surface of ethanol-water mixture (each $\times$ corresponds to a different $\%$ of ethanol in water) extracted from Lorenceau \textit{et al.} \cite{Lorenceau2020} and at the surface of sea water (each $\diamond$ corresponds to a different bubble size) extracted from Zheng \textit{et al.} \cite{zheng1983}}., 
  \label{fig:lifetimeNoSurfactants}
\end{figure}

Moreover, it appears that even so the bubbles look similar, their lifetime is not determined uniquely but follows some stochastic law. Studying the lifetime of bubbles quantitatively then implies to lead a statistical study\cite{haffner2018can}, all the more complex that many parameters affect the time evolution of the liquid film thickness on top of a bubble. 
Indeed drainage (section \ref{section:thinning}) and evaporation (section \ref{section:evaporation}) do not only depend  on the chemistry of the solutions but also on the bubble generation and the humidity of the room. 
Thus, studying lifetime of surface bubbles and comparing one system to another requires careful attention. 
Taking all precautions (control of humidity, and of bubble generation), Miguet \textit{et al.} \cite{Miguet2019} have shown two distinct behaviors of bubble lifetime distribution (figure \ref{fig:lifetimeNoSurfactants}.a). One set of bubbles generated in sucrose/water solutions (red curves) has its lifetime spreading from zero to about 30 seconds with a maximum around 8 seconds. 
 A second set of bubbles obtained in sweeteners/water solutions (green and blue curves) features a much more pronounced maximum around 23 seconds while random bursting events at earlier times, so presumably for thicker films, are still measured.
As the bubbles size, viscosity, density and surface tension of the solutions are quite similar for all baths, one would expect the same drainage laws.
Thus the first set of bubble appears to be "fragile" and to burst during film drainage for thickness much larger than 100 nm, whereas the second set of bubble presents a better resilience with respect to perturbations.
For instance, films as thick as 20 $\mu$m have been measured at the rupture in Mezcal solutions \cite{Rage2020}.

In the following, we will describe the state of the art for both families of bubbles.

\subsection{In absence of stabilising molecules}

In this part, we assume that no stabilising molecules are in the bath solution and thus that the bursting of the bubble occurs during film drainage and for thicknesses that might be larger than $100$ nm. 
In 1983, Zheng \textit{et al.} \cite{zheng1983} measure the lifetime of about one hundred bubbles at the surface of different water baths (tap, river, ocean), they found small variation of the average bubble lifetime (from 2.2 to 3.9 s) but the statistical features seem independent of the bath and seem to follow a Rayleigh distribution.

Actually, as the rupture of the film is a random failure event, the statistic law of lifetime which is a random positive variable $t_b$ can be described using a Weibull distribution, in which the probability $p$ to observe a time $t_b$ is given by 
\begin{equation}
p(t_b)= \beta/(t_b/t_{b0})^{\beta-1}\exp(-(t_b/t_{b0})^\beta) ,
\end{equation}
where $\beta>0$ is the shape parameter and $t_{b0}=<t_b>/\Gamma(1+3/4)$ is the scale parameter with $\Gamma$ the Gamma function and $<t_b>$ the average value of lifetimes.
If the rupture of the film is a rare event, possibly triggered by an external parameter, that occurs at any time independently of the film thickness and age, the lifetime of a bubble could take any value between zero and infinity and is expected to follow a Weibull distribution with an exponent $\beta=1$. Note that $\beta=2$ corresponds to the Rayleigh distribution previously mentioned. These different situations have been reviewed recently by Suja \textit{et al.} \cite{suja2020single} .

Recently, experimental studies with better statistics, concerning more than thousand bubbles \cite{Lhuissier2012,poulain2018,Lorenceau2020,Miguet2020,Gilet2007} have shown that the lifetime of bubbles at the surface of water (hot, deionized, river ... ) or mixture of water and alcohol is well described by Weibull distributions with  $\beta=4/3$. 
Surprisingly this result is quite robust (Figure \ref{fig:lifetimeNoSurfactants}(b)). 
The rather small number of bubbles observed by Zheng \textit{et al} \cite{zheng1983} may explain why their data are well described by a Rayleigh distribution.

 The exponent $\beta=4/3$ is supported by a model \cite{Lhuissier2012} that associates film lifetime to the rupture of the thin cells of the marginal regeneration due to the destabilization of the pinch, as described in section \ref{section:destabilisation}. 
 Perfectly bare interfaces indeed never exist and  even a small amount of impurities can create surface tension variations so that the pinch at the foot of the bubble can form.
 The authors introduce an \textit{ad hoc} parameter, the "efficiency of rupture" to calculate the average lifetime $<t_b>$. 
 Poulain \textit{et al.} \cite{poulain2018} completed this model and introduced the side-to-
side diffusion of impurities from one interface of the film to another.
Considering that rupture occurs if the diffusion time is smaller than the residence time of the contaminant in the cap, they propose 
\begin{equation}
  <t_b>=\alpha \tau_v Sc (R_{\text{film}}/\ell_c)^{1/2} ,  
\end{equation}
\noindent where $\tau_v=\eta \ell_c/ \gamma$ and $Sc=\frac{\eta}{\rho D}$, with $D$ the diffusion coefficient of the impurities in the aqueous solution, is the Schmidt parameter and $\alpha$ is a parameter that accounts for thinning or thickening of the film \textit{via} Marangoni effects. Note that $D \propto 1/\eta$ leads to $<t_b> \propto \eta^3$ 
which makes a very strong impact of the viscosity  on the bubble lifetime.
In the case of Mezcal and water-ethanol mixture, Rage \textit{et al.} \cite{Rage2020} show a dependence of the lifetime with alcohol concentration, with a maximum around 50 $\%$ that they correlate to a maximum of viscosity.
Lorenceau and Rouyer \cite{Lorenceau2020} have tested this scaling for water-ethanol mixture and found a relatively good agreement for the impact of ethanol concentration varying from 0 to 12 $\%$ in mass on the bubble lifetime (Figure \ref{fig:lifetimeNoSurfactants}-c).
This work has shown that the average lifetime of bubbles at the
surface of an ethanol-water bath increases considerably with the concentration of ethanol.
The ethanol concentration heterogeneities induced by evaporation, which in turn create upward Marangoni stresses induce a large value of $\alpha$ around 350 that is twice larger than the coefficient for sea water bubbles \cite{zheng1983}. As previously exposed in section \ref{section:evaporation} evaporation can be responsible for longer bubble lifetime.

\subsection{In presence of stabilising molecules}

In the presence of surfactants to stabilise the interfaces, different studies propose that the thinning time gives a good prediction for the bubble lifetime. 
This means that the stabilisation of the interfaces is efficient enough to avoid any rupture of thick film and to allow the film to thin until the end, \textit{i.e.} until its thickness reaches a few tens of nanometers.
In that case, the lifetime is expected to be fixed by the thinning rate and by the initial and final bubble cap thicknesses.
The initial bubble cap thickness does not impact dramatically the final lifetime since the drainage of a thick film is quite fast.
The exact value of the final thickness is still an open question \cite{rio2014thermodynamic}.
Most of the time, it is proposed to be around 100 nm because van der Waals interactions become important at such scales.
Nevertheless, it is well known that Newton black film, of thickness around a few nanometers, can exist and even be stable.
Thus, thin films do not necessarily rupture as soon as van der Waals forces become important and the whole story is not yet available.
Nevertheless, it has been shown \cite{Miguet2019} that the lifetime is not really sensitive to the final thickness value if it is chosen under $100$ nm.
Indeed, the evaporation leads to a thinning rate of a few tens of nanometers per second (Section \ref{section:evaporation}).
An incertitude on the final thickness of a few tens of nanometers thus leads to errors around a few seconds on the lifetime, which is small compared to the lifetime of a soap bubble.
This mechanism leads to a distribution of the bubble lifetime very different from what is observed in absence of surfactant. 
The distribution is actually much narrower and a typical shape is shown in Figure \ref{fig:lifetimeNoSurfactants}(a) in water in presence of sweeteners.
Unfortunately, no theoretical description leading to such a distribution is available in the literature.
In the following, we will concentrate on the prediction of the average bubble lifetime based on the thinning and on the comparison with the available experimental data.

\begin{figure}[!ht]
  \centering
    \includegraphics[width=1\linewidth]{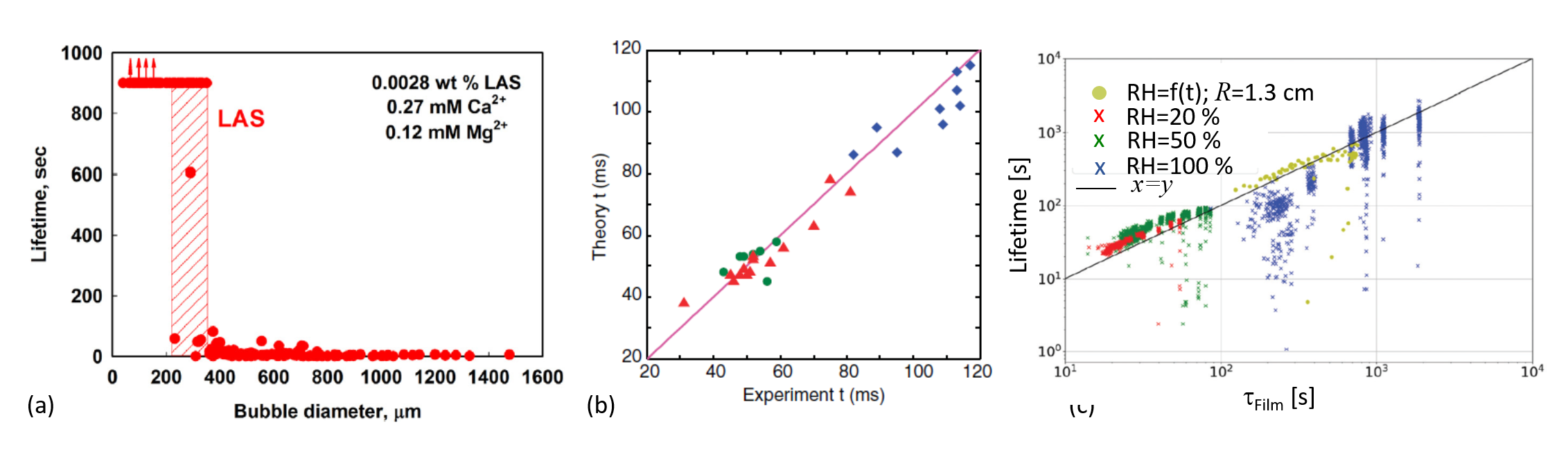}
  \caption{(a) Lifetime of microbubbles as a function of their diameter. An outstanding stability is observed for bubbles whose diameter is smaller than 200 $\mu$m. The figure is extracted from reference \cite{Pagureva2016} (b) Comparison between measured coalescence time between two bubbles and the Stokes-Reynolds-Young-Laplace model proposed by the authors of reference \cite{Vakarelski2010}, where the figure comes from. (c) Comparison between the measured lifetime of centimetric bubbles obtained for different atmospheric humidities and bubble size and the predicted lifetime obtained taking into account the film drainage and evaporation. Figure extracted from reference \cite{Miguet2019}.}
  \label{fig:lifetime}
\end{figure}

Let us first explore the influence of the bubble size.
As detailed in section \ref{section:thinning} and \ref{section:evaporation}, the thinning rate drastically depends on the bubble size.
Thus, this parameter is also expected to influence the bubble lifetime. 
In fizzy beverages, the bubbles are mostly quite sub-millimetric.
For a radius of $500$ $\mu$m, a density of 1000 kg/m$^3$ and a typical surface tension of $50$ mN/m, the Bond number defined in Eq. \ref{eq:BondNumber} is equal to 0.05.
In this case, the buoyancy is too small for the bubble to protrude and the surface area of the thin film is almost zero so that the probability for a hole to appear is very small. 
In such situations, a large stability of the bubbles can be observed, with static bubbles lasting hundreds of seconds \cite{Pagureva2016} (Figure \ref{fig:lifetime} (a)).
Their lifetime is  predicted by the hydrodynamic description of the bubble approaching the interface. 
Politova \textit{et al} propose that this velocity is given by an interpolation between the Stokes velocity, at which a sphere rises in an infinite bath and the Taylor velocity at which a sphere approaches a solid surface \cite{Politova2017}.
This model allows to predict the scaling of the lifetime of surface droplets with their radius up to 10 $\mu$m, which has every reasons to apply also for small surface bubbles. 
Nevertheless, as soon as the bubble impacts the surface with some non-zero velocity,  the precise dynamics of the bubble-surface approach is crucial to predict the bubble lifetime \cite{Vakarelski2010,Chan2011}. 
In this dynamic situation, a Stokes-Reynolds-Young-Laplace model taking into account the hydrodynamic-Laplace-disjoining pressure equilibrium describes the data very well \cite{Chan2011,Vakarelski2010} (see Figure \ref{fig:lifetime} (b)).

Pagureva \textit{et al.} \cite{Pagureva2016} also predict a slower increase of the bubble lifetime with the bubble size for higher bubble radius, which is reminiscent of the results obtained by different authors \cite{Miguet2019,Lhuissier2012,Poulain2018a}.
This is actually due to the fact that, although gravity forces are greater for big bubbles, the decrease of the capillary suction leads to longer lifetimes.
Nevertheless, for big bubbles, it is mandatory to take into account the influence of evaporation on the thinning to predict the right lifetime. 
This is true as soon as the Bond number is high enough for a thin film to appear. 
Indeed, as explained in section \ref{section:evaporation}, evaporation becomes important only in very thin films.
For surface bubbles submitted to evaporation and to the thinning through the marginal pinch described in section \ref{section:pinching}, Poulain \textit{et al} \cite{Poulain2018a} proposed that the bubble lifetime is given by \begin{equation}
\tau = 1.23 R \left(\frac{\eta}{\gamma}\right)^{2/5} \left(\frac{\rho}{J}\right)^{3/5}.
\end{equation}
\noindent This equation is obtained by integrating Eq. \ref{eq:LhuissierMassConservation} using the flow velocity given by Eq. \ref{eq:LhuissierScaling}.
For big bubbles, this lifetime actually overestimates the actual lifetime by up to 10 \% since an additional gravity driven drainage needs to be taken into account \cite{Miguet2019}.
Doing so, the experimental data are very well predicted by the thinning rate (Figure \ref{fig:lifetime}(c)).

Another parameter which is expected to influence drastically the bubble lifetime is the surfactants used to stabilise the interfaces \cite{Champougny2016,Pagureva2016}. 
Unfortunately, their influence is really mostly an open question. The models mentioned in the previous paragraph include the physical-chemistry only through the density, the viscosity and the evaporation rate.
Whereas both first parameters are not expected to vary a lot in presence of surfactants, the effect of surfactants on the third one is mostly unknown.
Nevertheless, the evaporation rate is expected to vary with the chemical potential and thus, with the surfactant concentration, which can vary along time due to the evaporation of water \cite{Mer2014}.
Additionally, as mentioned in section \ref{section:destabilisation}, the surface elasticity can affect the marginal regeneration and, thus the drainage. These are first leads to take into account the influence of the surfactants used to stabilise the bubbles.

\section{Conclusion and open questions}

To sum-up, we have presented the state of the art concerning the prediction of the lifetime of surface bubbles. 
In general, the bubble unstability is linked to two facts: (i) the bubble cap is constituted by a thin film, whose thickness decreases along time due to both drainage and evaporation and (ii) this thin film is unstable and eventually bursts.
We have shown that the current understanding is that two different behaviors exist depending on whether the film thins until its thickness reaches a few hundreds of nanometers or bursts at higher thicknesses. In the first case, determinist models that describe the thinning of the film down to a rupture thickness of the order of tenth to hundreds of nanometers perform correctely to calculate the bubbles lifetime. In the second case, the presence of a fatal impurity within the film and its propension to break it being a more random process, lifetime distributions are much more spread and only stochastic models may capture the physical mechanism(s) at play.
The scenario depends on whether or not surfactants are present to stabilise the thick film.

In absence of surfactants, the distribution of lifetimes is given by a Weibull distribution.
The bursting mechanism available in the literature involves the diffusion of impurities in the film, which cause the film rupture. 
Film thinning due to evaporation is likely to be rather negligible in such experiments since its impact is small on thick films.

In presence of surfactants, the film is expected to thin until its thickness reaches a few tens of nanometers.
The prediction of the bubble lifetime thus depends on our ability to predict the thinning rate of the film.
It is fixed by the evaporation and the drainage.
For tiny bubbles, no stable thin film appears and the evaporation is negligible. 
The lifetime is fixed by the approach velocity of the bubble to the bath.
For bigger bubbles, evaporation and drainage must be taken into account. 
The evaporation is a constant rate, which depends on external conditions such as atmospheric humidity, on the diffusion/convection ratio and on the chemical potential of the solution.
It has been shown that an accurate description of the evaporation rates necessitates to take into account the natural convection.

The drainage mechanism depends on the viscosity of the solution, on the bubble size and on the surfactants used to stabilise the interfaces. 
We have identified three main mechanisms.
For viscous bubbles, the cap is axisymmetric and the thickness decreases continuously from the bottom to the top of the bubble. The drainage is then expected to be exponential with time.
In presence of surfactants, a pinch is expected to appear in the vicinity of the meniscus, which slows down the drainage.
The destabilization of this pinch may lead to marginal regeneration, that in turn can affect the drainage.

Several references show that the drainage and evaporation rates are sufficient to predict the average lifetime of the surfaces bubbles in these different cases \cite{Politova2017,Poulain2018a,Miguet2019,Vakarelski2010}.

Many questions remain open and deserve to be addressed in a near future and we try to list some of them below. 

\begin{itemize}
    \item The mechanism at the origin of the eventual bursting of the film, whether they are thick (micrometers) or thin (tens of nanometers) is mostly unknown.
    \item The marginal regeneration phenomenon, the dynamics of the pinch, the origin of its destabilisation and its contribution to drainage are under current investigation.
    \item The impact of the chosen surfactants on bubble drainage and evaporation is crucial but remains an open question.
    \item There is still a lack of data concerning the distributions observed. Additionally, there is no theoretical prediction of the distribution in the presence of surfactants stabilising the interface.
    
\end{itemize}

\bibliographystyle{plain}
\bibliography{Biblio} 

\end{document}